# The Kids Are / Not / Sort of All Right[1]

Technology's Complex Role in Teen Wellbeing During COVID-19

CAROLINE PITT

The Information School, University of Washington, Seattle, Washington, USA, pittc@uw.edu

ARI HOCK

College of Education, University of Washington, Seattle, Washington, USA, arihock@uw.edu

LEILA ZELNICK

Department of Medicine, University of Washington, Seattle, Washington, USA, lzelnick@uw.edu

KATIE DAVIS

The Information School, University of Washington, Seattle, Washington, USA, kdavis78@uw.edu

We investigated changes in and factors affecting American adolescents' subjective wellbeing during the early months (April – August 2020) of the coronavirus pandemic in the United States. Twenty-one teens (14-19 years) participated in interviews at the start and end of the study and completed ecological momentary assessments three times per week between the interviews. There was an aggregate trend toward increased wellbeing, with considerable variation within and across participants. Teens reported greater reliance on networked technologies as their unstructured time increased during lockdown. Using multilevel growth modeling, we found that how much total time teens spent with technology had less bearing on daily fluctuations in wellbeing than the satisfaction and meaning they derived from their technology use. Ultimately, teens felt online communication could not replace face-to-face interactions. We conducted two follow-up participatory design sessions with nine teens to explore these insights in greater depth and reflect on general implications for design to support teens' meaningful technology experiences and wellbeing during disruptive life events.

**CCS CONCEPTS •Human-centered computing~Human computer interaction (HCI)~Empirical studies in HCI•Social and professional topics~User characteristics~Age~Adolescents**

---

[1] This title references a 2010 movie "The Kids Are All Right" exploring the lives of two teenaged siblings.

# 1 INTRODUCTION

The coronavirus pandemic raised concerns about potential negative impacts on young people's mental health and wellbeing [14,28,71]. School moved online for a large portion of youth in the United States, with mixed results; many families had to cope with illness and financial strain; and access to the daily activities and social interactions that support youth development and wellbeing were cancelled or severely reconfigured. Early reports suggest young people of all ages experienced stress, worry, unhappiness, and attention difficulties in the face of these challenging circumstances [12,43,67,91].

Networked technologies emerged as a lifeline for many youth during the pandemic, making it possible for them to complete schoolwork, maintain connections with family and friends, and fill the unstructured time that opened up in the midst of school closures and cancelled activities. Despite the evident value of having access to this lifeline (and we recognize that many youth lack such access), it is complicated by ongoing societal and scholarly discussions about technology's impact on youth wellbeing [32,35,40,63]. Within the research literature, some studies indicate negative effects, others show positive effects, and still others show little to no effect at all (see [61] for a review). Given this mixed research landscape, it is far from clear whether technology's increased role in youth's lives during the pandemic has ultimately been good, bad, or immaterial to their sense of wellbeing.

Although research on technology, youth, and wellbeing includes children from birth through emerging adulthood (e.g., [17,35,61,72]), we focus in this paper on adolescence because of the sensitive period it represents for mental health and wellbeing [60]. Adolescence also marks a time when technology use increases as parental monitoring diminishes [41,75]. Early evidence suggests that teens' use of technology increased considerably during the pandemic [12,67,91]. In this study, we pose the following research questions:

> RQ1: What trends do we see in subjective wellbeing within and across teens during the early months of the pandemic in the US?
> RQ2: In what ways are teens' interactions with technology during the pandemic associated with their wellbeing?
> RQ3: How can we approach the design of technologies that support teen wellbeing during crisis events?

We documented the daily activities, experiences, and feelings of 21 teens living in a metropolitan area in the Northwest United States over a three-month period to gain insight into technology's role in adolescents' sense of wellbeing during the early months (April – August 2020) of the coronavirus pandemic in the US. We conducted in-depth interviews with each participant at the beginning and end of the study and administered ecological momentary assessments (EMAs) three times per week for 10 weeks between the interviews. By the end of data collection, we had conducted 42 interviews and collected 581 completed EMAs. We used these data to examine daily changes in individual teens' wellbeing, technology's relationship to these changes, and other key factors that contributed to teen wellbeing. To explore themes from the interviews in greater depth, we conducted two participatory design sessions in which a subset of teens (*N* = 9) reflected on emergent themes from the empirical data and engaged in design exercises focused on making technology more meaningful and satisfying in their lives.

We found an aggregate trend toward increased wellbeing over the course of the study, but teens varied considerably in their individual, day-to-day levels of wellbeing. Participants described several factors that contributed to or undermined their wellbeing, including concerns about COVID-19, the transition to online school, increased time with family, prolonged physical separation from friends, and experiences with technology. Although technology use took place within a larger context of concerns and experiences, it nevertheless emerged as an impactful factor in teens' pandemic-related experiences. Teens reported an increase in their use of and reliance on networked technologies since the start of the pandemic, placing this change in the broader context of a considerable increase in unstructured, independent time. Using multilevel growth modeling, we found that how much total time teens spent with technology had less bearing on daily fluctuations in affective wellbeing than the satisfaction and meaning they derived from their technology use. Even when their experiences with technology were meaningful, teens felt that online communication could not, ultimately, replace face-to-face interactions with their friends. We explored these insights in greater depth during two follow-up participatory design sessions, during which teens reflected on the qualities of technology experiences that are more and less meaningful to them, as well as those that give rise to satisfying social interactions. We reflect on the general implications of these insights for design.

This work's primary contributions are:

1. Empirical evidence showing within- and across-person variation in American adolescents' subjective wellbeing during the early months of the coronavirus pandemic in the United States. This research is necessary to identify how teens coped amidst the global pandemic, as well as what factors support and undermine wellbeing in teens facing disruptive life events more generally.
2. Empirical evidence demonstrating technology's complex role in teens' experiences of subjective wellbeing during the pandemic. Our study joins an emerging and much needed body of work that uses longitudinal data to examine both between- and within-person variation in teens' technology use and wellbeing over time. To our knowledge, ours is the first of these studies to include a measure of meaningful technology experiences in addition to the more common measure of time spent with technology. Consequently, the current study contributes new insight into the kinds of technology experiences that teens find more or less satisfying (and why), and how those experiences relate to individual wellbeing.

## 2 RELATED WORK

### 2.1 Adolescent Mental Health and Wellbeing

Adolescence is a time of flux, a period of development accompanied by pubertal and neurobiological changes, increasing cognitive abilities, reconfiguring parent and peer relationships, and identity exploration [85]. Within this context of physical, cognitive, and social change, adolescents are considerably more vulnerable than younger children to experiencing mental health problems such as depression and anxiety [60]. Rates of depression spike in early adolescence and continue to rise throughout adolescence and into adulthood [11,85]. In fact, approximately half of lifetime mental health problems are established by age 14 [96]. The percentage of adolescents experiencing a major depressive episode (MDE) has been increasing in recent years [58]. In 2017, 13% of US adolescents (aged 12-17 years) reported an MDE in the last year [98].

The experience of stressful life events can contribute to mental health problems during adolescence [13,60,85]. The way adolescents cope with stressors is not always optimal, due to increased emotional reactivity and a decision-making system that is still developing [60,84]. Access to strong environmental supports—families, mentors, peers, schools, religious and civic groups—can help adolescents manage stress and develop resilience, as can a healthy lifestyle involving good nutrition, exercise, and sleep habits [60,96]. By the same token, the lack of adequate supports is likely to exacerbate the negative effects of stress. The emergence of a global pandemic in 2020 represents a stressful life event affecting adolescents worldwide. Although the pandemic has affected all ages, the vulnerability associated with adolescence warrants attention to its distinct effects on adolescents. Early research suggests that adolescents have indeed struggled with mental health challenges associated with the pandemic [43].

Despite adolescents' increased vulnerability to stressful life events, positive youth development scholars remind us that adolescence also represents a period of developmental opportunity, where, with the right support, youth can develop resilience, positive coping strategies, and emotion regulation skills that position them on a path toward emotional health as they enter adulthood [60]. Self-determination theory (SDT) describes how supports that meet youth's three basic psychological needs for *autonomy* (sense of initiative, acting in accordance with one's personal goals and values), *competence* (a feeling of mastery, opportunities for learning), and *relatedness* (a sense of belonging and connection to others) result in the promotion of wellbeing and self-motivation [22,77]. SDT addresses the dimension of wellbeing that relates to meaning and purpose, which are key to adolescent development and can act as buffers against stressful life events [20,30]. Other dimensions of wellbeing include experiences of pleasure and engagement [19,25,78].

Research has shown that pleasure, meaning, and engagement each support the two dimensions of subjective wellbeing: cognitive and affective wellbeing [25,79]. The former addresses the cognitive-judgmental dimension of wellbeing (one's assessment of overall life satisfaction), while the latter addresses the emotional aspect of wellbeing (how positive or negative one feels in the moment) [24]. In the current study, we examine both dimensions of adolescents' subjective wellbeing and the experiences—including technology-based ones—that affect them.

## 2.2 Technology and Adolescent Wellbeing

Research on the effects of technology on adolescent wellbeing is mixed (see [61] for a review). Some studies have found a negative relationship between technology use and indicators of adolescent mental health such as depression and anxiety (e.g., [6,10,89,90]); some have found positive effects (e.g., [21,74]); and still others have found little to no relationship between technology use and adolescent wellbeing [18,36,65].

One reason for these contradictory findings is the blunt methodological approach of largescale, cross-sectional surveys that look at aggregate-level trends across adolescents) [35,63]. The differential susceptibility to media effects model (DSMM) proposed by Valkenburg & Peter [92] illustrates why this approach can be misleading, masking the complexity of adolescents' relationships to the various devices and platforms they use daily. The DSMM posits that media (including interactive technologies) are used differently across youth and will have different impacts on individual youth depending on their individual dispositions, developmental level, and the social contextual factors associated with their media use. According to this model, studies that focus on aggregate-level effects will necessarily miss the person-specific uses and effects of technology.

Researchers who have begun to incorporate individual variation and longitudinal data into their study design (e.g., [2,6,18,36,64]) contribute to our emerging empirical understanding of the complexity associated with technology's impact on adolescent wellbeing. For instance, Beyens et al. [2] found considerable variation in the effects of social media use on momentary affective wellbeing across individual adolescents. By surveying adolescents six times daily for one week, the researchers found that passive social media use could have negative, neutral, or positive effects on wellbeing depending on the individual adolescent. The researchers also found differences in the relationship between social media use and wellbeing according to the type of platform used. This and other studies (e.g., [36,64,73]) demonstrate that variations in individual adolescents and the specific ways they interact with technology matter [35]. These studies also point to the value of gathering data from adolescents at multiple points across time in order to identify individual-level patterns in technology use and its relationship to wellbeing. Prior to this work, the vast majority of research on teens, technology, and wellbeing involved cross-sectional data [63]. Yet, even these newer studies still primarily focus on time spent using technology—particularly social media—rather than the *quality* of teens' time with technology. In the current study, we examine both time spent and teens' satisfaction with the technologies they use the most.

## 2.3 Designing for Adolescent Wellbeing

In the last decade, HCI researchers have become increasingly interested in designing for personal wellbeing, as evidenced by work in the areas of Positive Technology [76], Positive Computing [8], Positive Design [23], and Experience Design [29]. This work represents a movement away from an emphasis on usability and engagement in HCI and toward greater consideration for the promotion of meaningful activities that support and sustain wellbeing [37,46–48,54,68,88]. For example, Mekler and Hornbaek [54] focus on meaning as a core dimension of wellbeing and present a framework for the experience of meaning in human-computer interaction. In another recent example, Peters et al. [68] draw on the three basic psychological needs of self-determination theory (SDT)—autonomy, competence, and relatedness—[22] to present a model of wellbeing-supportive design, called Motivation, Engagement, and Thriving in User Experience (METUX).

HCI research has also begun to focus on how people use technology to support their wellbeing during stressful life events [34,50,55]. For instance, Liu et al. [45] investigated how chronically ill children used networked technologies to stay connected to their peers and maintain a sense of normalcy in their lives. Similarly, Maor and Mitchem [49] found that mobile technologies played an essential role in meeting hospitalized adolescents' learning, communication, and wellbeing needs. In another example, Iacovides et al. [34] examined the role of gaming during difficult life experiences and found that games gave participants a needed break from the stress at hand, helped them to work through their feelings about the current situation, and provided them with social connection.

Building on such work, researchers have begun to explore the potential of games to support adolescent mental health [15,16,53,62]. Virtual reality (VR) has also been a focus of recent design work to support adolescent wellbeing [5], as well as pediatric patient support [44] and investigations of wellbeing more generally [9]. With respect to designing for wellbeing in the specific context of COVID-19, Riva et al. [76] draw on the Positive Technology framework to identify technological strategies to support wellbeing during COVID-19. They identify VR, m-Health tools, video games, and social exergames as specific types of technologies that can be designed to reduce stress and increase positive feelings (VR, m-Health), induce feelings of flow and awe (VR), and support connectedness among individuals and groups of people (social VR, m-Health, video games, and social exergames). In the current work, we investigate the types of interactive experiences that adolescents turned to during the pandemic to support their wellbeing, and we reflect on how their experiences can inform the design of future technologies to support adolescent mental health.

## 3 METHOD

The current study answers growing calls to apply methods that uncover rather than gloss over individual variation in young people's technology use and its impact on their wellbeing [2,35,64,92]. We used in-depth interviews and ecological momentary assessments (EMAs) in combination to examine the individual experiences of teens during the early months of the coronavirus pandemic (April—August, 2020) in the United States. This mixed-method approach provided insight into aggregate and individual trends in teen wellbeing, daily experiences affecting wellbeing, and the particular role played by teens' interactions with technology. Engaging a subset of teens in follow-up participatory design sessions provided the opportunity to share emerging themes with them, further refine and explore those themes, and use the resulting insights to identify design implications for supporting teens' positive technology experiences and subjective wellbeing during disruptive life events.

### 3.1 Participants

We recruited 22 teens from a metropolitan area of the Northwest United States. One participant dropped out of the study during data collection, resulting in a total of 21 teens aged 14-19 years ($M$ = 15.8 years). Inclusion criteria for participation in the study were age (13-19 years), residence in the same metropolitan area, and at least somewhat reliable internet access. Approximately half of the teens ($N$ = 11) were recruited from an afterschool science program composed of youth from diverse racial, socioeconomic, and cultural backgrounds. The remaining 10 teens were recruited through Twitter and word of mouth in the first and last authors' social and professional networks. Table 1 shows the demographic breakdown of participants. All participants received financial compensation for their participation in the study.

Table 1: Participant demographics

| Participant Demographic | Percentage ($N$ = 21) |
|---|---|
| **Age (years)** | 14yo: 19%, 15yo: 28.6%, 16yo: 19%, 17yo: 23.8%, 18yo: 4.8%, 19yo: 4.8% |
| **Grade (U.S.)** | 8th: 4.8%, 9th: 23.8%) 10th: 23.8%, 11th: 19.4%, 12th: 14.3%, First year university: 4.8% |
| **Race/Ethnicity** | White: 42.8%, Asian: 19%, Hispanic or Latina: 9.5%, Black: 4.8%, Other/mixed ethnicity: 23.8% |
| **Gender** | Female: 47.6%, Male: 47.6%, Non-binary: 4.8% |

### 3.2 Data Collection Procedures

Data collection started on April 15, 2020 and ended on August 27, 2020. At the start of the study (April 15), both public and private schools in the area had been closed since March 12, and a stay-at-home order had been in place since March 23. In early May, the area moved into Phase 1 of a reopening plan, which allowed for some outdoor recreation but no social gatherings, and the stay-at-home order remained in effect. In mid-June, the area moved to Phase 2, which allowed for socially distanced gatherings of no more than five people per week. The study ended in Phase 2.

The three researchers (first, second, and fourth authors) who interacted directly with the teen participants were experienced in conducting interviews and participatory design sessions with adolescents. All study procedures were approved by our university's Institutional Review Board.

#### 3.2.1 Semi-structured interviews

Each teen participated in two semi-structured interviews via video conferencing approximately 12 weeks apart. Interviews ranged in length from 43 – 118 minutes. The interview protocols included questions about participants' daily activities, technology use, family and peer interactions, school experiences, sleep patterns, news consumption, and feelings and general reflections about the unfolding coronavirus pandemic and its impact on their lives. The second interview included questions inviting participants to reflect on changes and consistencies in the aforementioned topic areas. In this interview, we also displayed several visualizations of participants' daily responses to EMA questions about their technology use and wellbeing, inviting them to reflect on the patterns they observed (see 3.2.2). In these questions, we asked teens to describe their reasons for rating certain app experiences (e.g., with Instagram or Netflix) as more or less satisfying, providing insight into how participants used different applications, how they defined satisfaction, and how their definitions related to wellbeing. All interviews were audio recorded and transcribed verbatim.

*3.2.2 Ecological momentary assessments*

Between the two interviews, we administered ecological momentary assessments (EMAs) to participants via their mobile phone three times per week for 10 weeks. EMAs enable the gathering of intensive, in-situ data across time, increasing recall of daily experiences in comparison to retrospective accounts and enabling researchers to track variation within and across individuals over time [61,80,86]. During the 10-week period, we alternated the days of the week on which we administered surveys in order to gain as full a perspective as possible on teens' daily experiences and feelings. Participants received daily surveys at approximately 11am, a time by which most of them were awake (interview data determined that sleep patterns were varied) and early enough for them to recall their activities from the previous day. The overall compliance rate was very high: 96.8%.

The EMA questions mirrored the questions asked in the interview and were intended to track participants' in-situ experiences and wellbeing levels over time. The technology-related questions asked participants to report the amount of total time they had spent with technology during the previous day, time spent on individual applications (e.g., Instagram, YouTube, Zoom), and satisfaction ratings for each app experience (0 = not at all satisfying; 100 = very satisfying). Following Verduyn et al. [93], we measured affective wellbeing with the question "How do you feel right now?" (0 = very negative; 100 = very positive). In addition, we asked participants how lonely, worried, and connected to others they felt, all dimensions of subjective wellbeing [38,79]. To measure cognitive wellbeing, we administered the Satisfaction with Life Scale [24] at the start and end of the study. Possible scores range from 5 (extremely dissatisfied) to 35 (extremely satisfied) [66]. We used the second interview (3.2.1) to confirm the accuracy of participants' EMA responses and to give them the opportunity to explain the reasoning behind their answers.

*3.2.3 Participatory design sessions*

At the conclusion of the interview and EMA data collection sequence, we invited all teens to a two-hour participatory design session via video conferencing. This approach provided an opportunity to explore in greater depth major technology-related themes that surfaced in the interviews and EMAs. Seven US-based teens participated (3 in session one, 4 in session two), as well as two additional teens from a concurrent study taking place in Germany (session two). Each design session focused on a specific theme related to technology and wellbeing that emerged from the interviews and EMAs. The first session focused on how to avoid the feeling of wasted time with technology, and the second focused on enhancing the meaningfulness of online social interactions. We started each session with exemplary quotes from the interviews, inviting teens to discuss their ideas surrounding these topics with each other and generate design ideas for addressing the issues. The quotes were displayed on Padlet, an interactive website allowing users to add notes and drawings to a digital sticky note wall. Drawing on best practices from design work with children and teens [3,4,27,94,95], each session progressed through a series of idea-generation questions using digital sticky notes, which prompted the co-designers to discuss and remix each other's ideas. We recorded both sessions, wrote analytical memos, and preserved the digital artifacts, including images of the sticky wall and exports of the text.

### 3.3 Data Analysis

*3.3.1 Interview analysis*

We used thematic analysis to analyze the interview transcripts [7], adopting both emic and etic approaches [51] to identify themes that were emergent in the data (emic) and related to our research questions and related work (etic). The last author reviewed five transcripts to generate an initial "start list" of codes [57], which were then shared and refined through discussion with the first and second authors. The final coding scheme (see supplementary material) captured themes related to sources of stress, experiences of loss, and supports for wellbeing. Technology-related codes captured themes related to teens' patterns of technology use and their reflections on the positive and negative dimensions of technology in their lives. Adopting a consensus coding approach [42,82], the three researchers individually coded three transcripts and came to consensus on appropriate code application. They divided the remaining transcripts and entered their coding and narrative memos into Dedoose. Through frequent discussion, the researchers ensured that codes were applied accurately and consistently. We shared emergent themes related to technology with participants of the two participatory design sessions as a form of member check on the validity of our findings [42,51].

*3.3.2 EMA analysis*

We defined a set of *a priori* questions for quantitative analysis, drawn from the first two research questions guiding the study and emergent themes from the interview analysis. We used these questions to determine both descriptive and inferential statistical procedures. The descriptive procedures included generating individual growth trajectories of the wellbeing measures (how do you feel, how connected, how worried) for each participant, as well as calculating individual means, standard deviations, and ranges for each participant on each measure [81]. The inferential procedures included a paired-samples *t*-test to determine average change in life satisfaction between the start and end of the study. We fit a series of linear mixed models to determine whether aspects of teens' technology use (time spent and satisfaction with technology) were associated with changes in their affective wellbeing over time. The models included random intercepts to account for within-person correlation and random slopes to allow for heterogeneity of association between persons [39]. In each model, we controlled for age and gender due to prior work indicating differences in technology use based on these demographic categories [74]. An anonymized dataset is available for researchers who wish to replicate our analyses (see supplementary material). All quantitative analyses were conducted in R version 6.3.2.

*3.3.3 Design session analysis*

We generated and triangulated salient themes using the session recordings, researchers' analytical memos/notes, and design artifacts [1,97]. The research team (first, second, and fourth authors) used these sources as the basis for a series of collaborative discussions about the way participants responded to and elaborated on the focal themes in their sessions. Through these discussions, we documented the degree to which the themes resonated with participants and the various ways they extended the themes through the design exercises. We used the resulting insights in conjunction with relevant prior work (e.g., [31,54]) to reflect on implications for the design of meaningful technology interactions.

## 4 RESULTS

Our interviews revealed that the pandemic had affected nearly every aspect of teens' lives—school, family, friendships, organized activities, leisure time, technology use, exercise, and sleep. A detailed account of the changes in each of these domains and how they affected teens' wellbeing is far beyond the scope of the current study. Our primary purpose in this paper is to examine technology's relationship to teen wellbeing during the pandemic. At the same time, focusing exclusively on technology-based experiences overlooks the broader context of teens' lives that inform variations in their sense of wellbeing, including the distinct role played by technology [70,83,87]. Therefore, as we report the results of teens' subjective wellbeing, we draw on data from the interviews to illustrate the many non-technological factors they discussed in relation to how they felt on any given day during the pandemic. We use participant identification numbers (with the prefix SP—study participant) to identify direct quotes rather than names to protect participants' privacy.

### 4.1 Variation and Change in Teen Wellbeing

*4.1.1 Overall movement in a positive direction*

Teens started the study (April 2020) with considerable variability in their self-reported satisfaction with life (cognitive wellbeing), ranging from dissatisfied (3 teens) to very highly satisfied (2 teens) [66]. A paired-samples *t*-test showed an aggregate-level trend toward increased cognitive wellbeing between the beginning and end of the study ($t(18)$ = 2.939, *p* = 0.0087), though there was considerable variation in life satisfaction between participants at each timepoint ($M_{Time1}$ = 22.47, sd = 5.96; $M_{Time2}$ = 25.58, sd = 5.12).

A comparison between teens' descriptions of their daily experiences at the time of their first and second interviews surfaced notable changes that likely contributed to this general trend. In their first interview, all teens spoke about common adolescent experiences and activities that had either been cancelled or severely pared down, such as in-person school, sports practices and competitions, music rehearsals and trips, college campus visits, afterschool and summer jobs, driver's education classes, and high school graduation. These widespread cancellations left many teens feeling disoriented by the lack of structure in their lives and without a clear purpose. SP05 (boy, age 18, 12th grade) reflected: "I don't have anything I'm aiming towards or have to do, so it's kinda just…I'm sitting idle."

By their second interview, many teens had resumed some of the activities that had been cancelled at the start of the pandemic, bringing back some of the familiar patterns from their pre-pandemic lives. SP22 (girl, age 15, 10th grade) was able to play volleyball again, albeit in a restricted, socially distanced format. SP02 (girl, age 19, college freshman) and SP17 (boy, age 17, 12th grade) had found hourly jobs. SP11 (girl, age 16, 10th grade) got her driver's license and was able to get together with her friends in person. SP20 (girl, age 17, 12th grade) also spoke about meeting up with friends for socially distanced picnics since the weather had improved and state restrictions on public gatherings had eased. In light of the fact that not being able to see their friends was the most frequently cited pandemic-related challenge in the first interviews, being able to see friends again was particularly welcomed by those teens who had done so by their second interview.

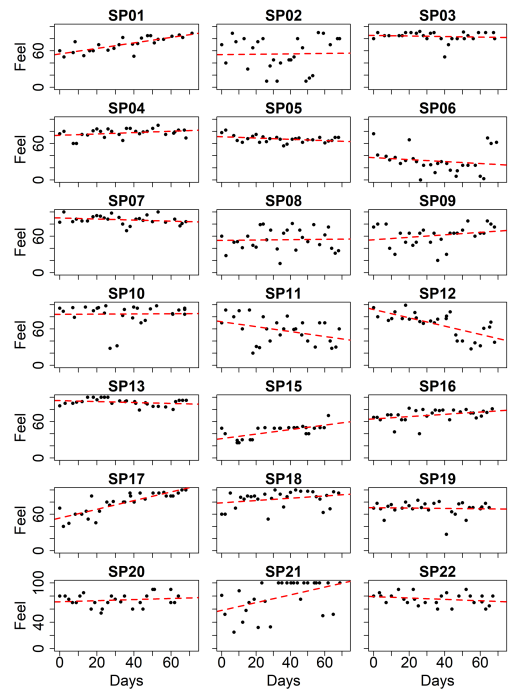

Figure 1. Individual plots of participants' daily responses to "How do you feel right now?" (0 = very negative; 100 = very positive) over the 10 weeks of EMA surveys. Linear trend lines are overlaid to summarize general trends.

### 4.1.2 Fluctuating feelings within and across teens

As seen in Figure 1, teens varied considerably in how positive they felt on any given day during a 10-week period between April and August 2020. We saw similar within- and across-participant variation in how connected and worried teens felt throughout the 10-week period (see supplementary material). Table 2 shows the results of regressing *Feel* on time for each participant in our sample, as well as the means, standard deviations, and ranges for individual participants. (Equivalent statistics for how connected and worried teens felt are available in the supplementary materials.) These data illustrate the variation in wellbeing across teens (differences in means and ranges), as well as the variation within individual teens over time (several teens had large standard deviations and ranges). A majority of teens did not display a statistically significant trend in either a positive or negative direction for any of the wellbeing measures. Of those teens who did show a trend, more teens moved towards increased wellbeing (increased positive feelings and connectedness, decreased worry), mirroring the results reported in 4.1.1. We showed participants their individual growth charts in the second interview and invited them to reflect on the patterns they saw. Below, we describe the most salient aspects of teens' lived experiences that affected how they were feeling from day to day.

Table 2. Summary statistics (Mean, (standard deviation), range), slope estimates, and 95% confidence intervals (CI) derived from regressing *Feel* (How do you feel right now?) on time for each participant ($N$ = 21).

| ID | $M$ (sd) range | Estimate (CI) |
| --- | --- | --- |
| **SP01** | 71.1 (12.1) 50-89 | 0.45 (0.34, 0.57)* |
| **SP02** | 54.9 (26.9) 10-90 | 0.03 (-0.45, 0.51) |
| **SP03** | 83.7 (8.5) 50-92 | -0.05 (-0.15, 0.06) |

| ID | M (sd) range | Estimate (CI) |
|---|---|---|
| SP04 | 77.4 (7.3) 60-90 | 0.11 (-0.02, 0.25) |
| SP05 | 67.5 (5.6) 56-82 | -0.11 (-0.21, 0)* |
| SP06 | 31.0 (20.1) 0-76 | -0.16 (-0.63, 0.3) |
| SP07 | 87.1 (7.2) 69-100 | -0.09 (-0.2, 0.03) |
| SP08 | 54.1 (17.6) 15-81 | 0.03 (-0.26, 0.32) |
| SP09 | 61.3 (16.8) 20-85 | 0.2 (-0.1, 0.51) |
| SP10 | 84.6 (17.4) 28-98 | 0.02 (-0.15, 0.19) |
| SP11 | 57.5 (21.1) 20-91 | -0.41 (-0.74, -0.07)* |
| SP12 | 67.7 (19.7) 27-99 | -0.68 (-0.94, -0.43)* |
| SP13 | 91.9 (5.8) 79-100 | -0.08 (-0.18, 0.03) |
| SP15 | 44.0 (10.9) 25-70 | 0.37 (0.16, 0.58)* |
| SP16 | 70.8 (10.1) 40-82 | 0.19 (0.07, 0.31)* |
| SP17 | 78.8 (17.7) 40-100 | 0.72 (0.54, 0.89)* |
| SP18 | 85.3 (13.5) 52-100 | 0.19 (-0.05, 0.43) |
| SP19 | 69.4 (11.7) 27-83 | -0.03 (-0.17, 0.11) |
| SP20 | 73.8 (9.9) 54-90 | 0.08 (-0.09, 0.26) |
| SP21 | 78.7 (25.8) 25-100 | 0.59 (0.15, 1.03)* |
| SP22 | 75.4 (8.9) 60-90 | -0.1 (-0.24, 0.03) |

* Indicates change over time in a statistically significant direction ($p < .05$).

*COVID-19.* Many teens observed higher-than-usual levels of worry and pointed to the pandemic as the primary reason. These teens were concerned about uncertain futures, including the possibility of themselves or their family members getting sick. SP02 (girl, age 19, college freshman) lived alone with her mother, who did late-night shift-work in a factory. She reflected: "I worry about my mom constantly, especially when she goes out…I don't know what I would do like if we did have like a scare or something like that." SP03's (girl, age 17, 11th grade) worry became a reality when her father fell ill with COVID-19 (thankfully, he made a full recovery). Her mother was an essential worker and could not risk getting sick, so it fell to SP03 to care for her father by giving him his medication and monitoring his breathing. For SP04 (girl, age 16, 10th grade), her worry was more generalized: "I think the biggest challenge is that no one knows the future. No one knows what's going on, so I can't ask an adult for… I can't be like, 'When's this going to end, what's going to happen?'" SP04 noted that her stress was magnified by encountering regular updates and discussions about the pandemic on social media, where she also saw stressful news about the Black Lives Matter protests and the US presidential election campaigns.

*School.* Next to COVID-19, school was the second most frequently-cited source of worry for teens. Although the majority of teens said they had less work overall—something they appreciated—the work that they were assigned and the way online classes were managed created new sources of stress. Especially for teens attending public school, work was often assigned at the start of the week for students to complete at their own pace. SP22 (girl, age 15, 10th grade) reflected on why this arrangement did not work for her: "…having to try to figure out a lot of stuff by myself has been a huge struggle with online school….I think the biggest is just there's no hands-on, face-to-face interaction, which is really, really hard."

*Family.* Family emerged as both a source of comfort and a source of stress for the teens in our sample. SP13 (boy, age 14, 9th grade), who consistently reported high levels of wellbeing throughout the study (see Figure 1), characterized increased family time as a distinct benefit of the pandemic: "I feel a little bit more boost in self-esteem and generally more happy being around family." In contrast, other teens described existing family tensions that were magnified by everyone living in close proximity during lockdown. SP11 (girl, age 16, 10th grade), who reported feeling increasingly negative during the study (see Figure 1), reflected: "I think me and my mom have always clashed. As much as I appreciate her being home a lot to hang out with me, I think that me and her have never been able to actually hang out and not fight for a whole day."

*Friends.* In addition to family, friends played an important role in how teens felt from day to day. Teens were grateful that they could maintain some form of connection with their friends online, noting that it helped them to feel connected. Still, they mourned not being able to hang out with their friends in person. SP22 reflected: "I think honestly, just not being able to go out and socialize and stuff has been the most challenging part, just 'cause I'm missing my friends all the time." Extroverted teens seemed to experience the greatest sense of loss, as SP21 (nonbinary, age 14, 8th grade) explained: "I get a certain energy out of seeing my friends because I'm an extrovert, and so it kind of hurts to not be able to see my friends…"

*Self-disposition.* Teens varied in the general attitude they brought to the pandemic, with some able to adapt to the changes in their lives while others struggled. SP13 reflected: "The first couple days I was kind of shocked about what was going on, and then I kind of adapted to it and kept on rolling on." SP15 (boy, age 16, 10th grade) said that he had never felt much stress related to the pandemic, explaining: "I'm kind of a stress-free person…A lot of the time is pretty much as stress free as [it] can be…" This stance stood in stark contrast to SP21, who experienced anxiety and depression before the pandemic: "COVID has drastically changed who I am as a person, who I wanna be as a person, how I'm going to take care of myself. All of that. And so that's really, really intense for me."

**4.2 When Time Opens Up and Life Moves Online**

*4.2.1 Free time and time sinks*

The cancellations and pared-down activities resulting from the pandemic left teens with a lot of extra, unstructured time. Whether teens felt good or not about their use of time often related to how they approached their use of networked technologies. Among those teens who offered positive assessments, they typically described using technology to pursue new or existing interests. For instance, SP01 (boy, age 15, 10th grade) was inspired by seeing examples of video-game editing on Instagram and decided to develop his own video-editing skills. SP12 (boy, age 15, 9th grade) was using his newfound time to work on his computer programming and geocaching badges for Boy Scouts. Siblings SP04 (girl, age 16, 10th grade) and SP07 (boy, age 14, 9th grade) were deepening their knowledge of music in a variety of on- and offline ways. SP04 was participating in an online songwriter's lab and practicing her ukulele. SP07 was exploring orchestration and reading about music composition.

Networked technologies were equally central to those teens who felt less positive about how they were spending their time. Although all teens valued having their devices to keep them connected and entertained, many of them were critical of how and how much they were using technology. SP09 (girl, age 15, 9th grade) reflected: "I kind of wish I wasn't using my phone as much, just doing stupid things. I'll be re-watching old TV shows or playing dumb games, and I don't really know what else to do." Like SP09, several teens struggled to find ways of using technology that left them feeling satisfied rather than wondering what they had to show for their day.

*4.2.2 Time spent vs. time well spent with technology*

Our analysis of the EMA data revealed that teens' levels of satisfaction with their technology use related directly to their daily reports of affective wellbeing and feelings of connection to others. We asked teens to list their four most frequently used applications during the previous day and rate how satisfying their experiences were with each one (0 = not at all satisfying; 100 = extremely satisfying). We used the average of these four ratings as the primary independent variable in two linear mixed models predicting changes in affective wellbeing (measured with the question "How do you feel right now?" 0 = very negative; 100 = very positive) and connectedness to others (0 = not at all connected; 100 = very connected). Controlling for gender and age, we found that each 10-point increment of the average level of technology satisfaction reported was associated with a 3.8-point difference in teens' affective wellbeing (95% CI: 2.3, 5.8; $p < 0.0001$) and a 4.0-point difference in teens' feelings of connectedness to others (95% CI: 2.2, 6.1; $p < 0.0001$). Notably, in two separate linear mixed models, neither total time spent with technology (95% CI; -0.9, 0.9; $p = 0.94$) nor total time spent on social media (95% CI: -1.6, 2.6; $p = 0.59$) predicted statistically significant differences in affective wellbeing, controlling for gender and age. (See supplementary materials for tables of all fitted models.) These results indicate that how satisfied teens felt with their technology use mattered more to their sense of wellbeing than how much time they spent with technology.

In their second interviews, we asked teens to reflect on how they approached rating their technology use as more or less satisfying. Teens consistently reported that they reflected on how good they felt after using a particular application. What made them feel good varied across platform and according to their purpose for using it. On platforms like Netflix, YouTube, and TikTok, teens typically considered how much they enjoyed the content they had watched. Several teens observed that even if they had started off enjoying the content, they sometimes ended a session feeling dissatisfied because they felt they had watched for too long. For instance, SP05 (boy, age 18, 12[th] grade) rated YouTube sessions as being least satisfying when he found himself watching a series of videos that played automatically after the video he had gone to YouTube intentionally to watch: "I would only be watching a couple of videos a day if it weren't for all the random stuff I watch in between all those [intentionally chosen] videos."

Another factor that many teens considered was how much they had learned while using a particular platform. SP20 (girl, age 17, 12[th] grade) explained why she consistently rated Instagram as highly satisfying: "Instagram keeps me engaged and shows me things I didn't know about." Related to the satisfaction of learning something new was the feeling of accomplishing something challenging, which was mentioned most often by teens who played video games or pursued their interests like music composition, video editing, and merit badges. SP08 (boy, age 16, 11[th] grade) reflected on the satisfaction he experienced from playing Runescape and earning XP (experience points) as he moved to higher and higher levels. He explained: "I want to be able to feel like I accomplished something that day. And that feeling of getting that one level, even though it's online, it's…pixels all put together, it feels very accomplishing and it's kind of very satisfying."

Whether using video chat, private messaging, or social media, teens often gave high satisfaction ratings to the time they spent online with friends. For instance, SP03 (girl, age 17, 11[th] grade) explained why she gave her highest satisfaction ratings to FaceTime conversations with her friends: "It's just like I miss them so much so like, any time I get to FaceTime or get to see their faces, it's satisfying to me. It makes me happy." Not all interactions with friends were equally satisfying, however; teens sometimes gave lower satisfaction ratings to online conversations with friends if they had argued or did not have anything to talk about.

*4.2.3   Facetime is not face-to-face*

Teens typically pointed to communicating with friends as the most valued function of networked technologies during the pandemic. Schoolwork, entertainment, and pursuing interests were also highly valued, but talking to friends helped address the deep sense of interpersonal loss that many teens were experiencing during lockdown. At the same time, all teen participants observed that, ultimately, technology was not an adequate substitute for in-person interaction.

Teens explained the inadequacy of online peer interactions by pointing to the fact that their pre-pandemic online interactions were fundamentally grounded in their shared, in-person experiences. For many teens, a large portion of their online communications before the pandemic revolved around talking about what was going on at school and in other aspects of their shared offline lives or making plans to meet up in person. Without the ability to hang out in person, there were fewer reasons to reach out online. SP05 (boy, age 18, 12th grade) noted the lack of conversation topics since the start of the pandemic: "[Communication with friends] is way less frequent because there's just less to talk about, I guess. Before coronavirus, we would talk about homework and what's going on at school…the content just isn't there for a good, nice conversation." To address the challenge of having nothing to talk about online, some teens turned to games or game-like experiences. For instance, SP16 (girl, age 15, 10th grade) and her friends started using House Party, which allowed them to play games like trivia and Apples to Apples during their video calls. SP02 (girl, age 19, college freshman) started playing Minecraft with her friends.

Several teens directly compared online and offline communication to explain why the former often felt less satisfying to them than the latter. SP02 noted the energy that results from sharing the same physical space and witnessing the same events as they unfold: "… that energy that we have when we're meeting in real life isn't the same as when you're texting someone." SP22 (girl, age 15, 10th grade) reflected on the challenges presented by reduced social cues online: "For me, it's hard to just read emotions and stuff over text, and I'd much rather just see [my friends] in person." These quotes address the visceral quality of in-person communication that teens missed since moving the bulk of their peer interactions online.

### 4.3 Going Deeper with Design

#### 4.3.1 Time (with technology) well spent

The first design session drew out specific aspects of the ideas surrounding meaningful technology that surfaced in the interviews and EMAs (4.2.1 and 4.2.2). Participants identified with the interview quotes we shared with them, reflecting on how easy it was to sink time into technology use without meaning to do so. One participant wrote on the Padlet: "Sometimes when I'm on my phone time flies by so fast and I don't realize that until a lot of time has went by until after. Then I end up regretting wasting my time on doing nothing when I could've done something fun." In addition to the distraction of technology, participants also mentioned that the removal of structure made them feel somewhat adrift. One sticky note stated: "Personally, I work well when I have lots of assignments and when I'm in a higher-stress environment. Now that I don't have many deadlines or important tasks, I find it harder to stay motivated."

The group of teens discussed ways of addressing these issues, which led to ideating features of an application or tool that would help with scheduling and time management. There was a strong focus on providing some structure while avoiding rigidity and adjusting to personal needs. One sticky note mentioned: "having cushions [of] time so if you mess up or it takes longer than expected you don't feel stressed out." Another strategy for providing time-management support—as opposed to punishment—involved incorporating social aspects into the technology. One note read: "It would be cool if I can share my reminders with someone so even if I wouldn't want to do that I can have someone do it with me or remind me." The teens also proposed features that would help users engage in personally meaningful activities during their downtime while avoiding the feeling of wasted time. One sticky said: "It would be cool if my calendar app gave suggestions of activities for when I have free time." This idea, endorsed by all participants, illustrates how teens were not simply interested in tools to help them get work done; they also wanted support finding value in their non-work uses of technology.

*4.3.2 Meaningful (virtual) connections*

Drawing on themes from 4.2.3, the second design session focused on deepening connections with friends and acquaintances during social distancing restrictions and other complicating factors of the pandemic. During the session, participants discussed the different dimensions of connection difficulty, from not having as much to talk about with close friends to not interacting with most of one's classmates at all. Mirroring themes from the interviews, one sticky note said:

> "[…] when I'm at school everyday, there's something new that I also experience everyday which then makes it easier to have a conversation with my friends versus me being at home, I noticed that my friends and I would have less to talk about over facetime because we were both home so often."

Participants also reflected on how the pandemic had severely curtailed interactions with acquaintances. As SP09 (girl, age 15, 9th grade) explained "…other than your like close friends, you kind of have fallen out of contact with a lot of other people, other than like seeing their Instagram posts. […] It's harder to stay in contact when there's not school you're seeing and interacting with people all the time."

Participants ideated strategies to increase connectedness, such as playing games, watching movies, and texting at regularly rescheduled times. Through their discussion, they converged on design ideas to facilitate a sense of co-presence, shared experience, and conversation topics with both close friends and acquaintances. The co-designers turned next to developing ideas for a system "[…] that gives suggestions on what a group of people would like to do together based on each individual's interests. A virtual reality setup where you can meet up with friends and go on trips together." VR was identified as a promising technology for generating a feeling of shared experience. Another note elaborated:

> "[the VR system] would allow friends to meet up in locations around the world. You could input some of your hobbies and favorite places and it could suggest places to go and things to do with different friends. You could also just let the technology pick for you and go somewhere [in VR] without knowing where."

Participants also suggested that the system have a variety of multiplayer games and that users would be able to specify which populations they wanted to be able to interact with (friends, acquaintances, classmates, the neighborhood, and so on). Participants noted that such a system could help to expand the diminished social worlds they were experiencing during the pandemic. The session concluded with co-designers detailing the features and sketching their designs, including the various options for VR experiences, such as a beach, movie theater, or outdoor Parisian cafe (see Figure 2). These design concepts reinforced the idea that teen participants were feeling the lack of in-person interaction with their friends and trying to find the means to forge stronger social connections, yet finding the current options an insufficient substitute to being physically present.

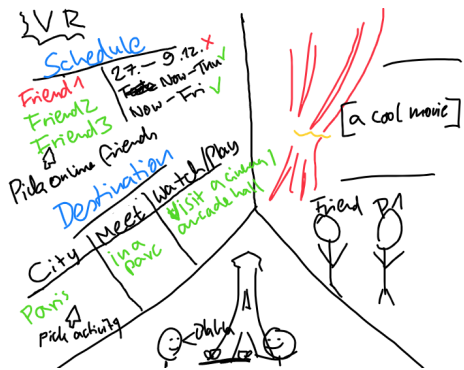

Figure 2. A design session participant's digital sketch of a social VR experience.

## 5 DISCUSSION

The current study contributes new empirical insight into teens' wellbeing and technology use during the early months (April – August 2020) of the coronavirus pandemic in the United States. Pairing in-depth interviews with daily ecological momentary assessments (EMAs), we were able to identify changes in subjective wellbeing within and across teens over time; key factors that played a role in these changes; and the specific role of teens' daily technology use. Although many factors affected teens' wellbeing during this period, including the transition to online school, prolonged separation from friends, and direct experiences with or concerns about COVID-related illness, we found that technology played a meaningful role and shaped other aspects of teens' experiences (e.g., as a lifeline to friends). A major contribution of this study is empirical evidence describing the kinds of technology experiences that teens find more or less satisfying, as well as evidence demonstrating that the satisfaction and meaning that teens derive from their technology use is more important to their daily levels of subjective wellbeing than the amount of time they spend using technology. To the best of our knowledge, ours is the first study to incorporate a measure of meaningful technology use in a longitudinal investigation of the connection between teens' technology use and wellbeing (see [2,63]).

By inviting teens to reflect on their reasons for rating technology experiences as more or less satisfying, we were able to identify key characteristics associated with teens' meaningful technology use. Several of these characteristics align with prior work exploring meaning in human-computer interactions, such as the feeling of wasted time when scrolling purposelessly online and the value of connecting with others and accomplishing something effortful (e.g., [54,88]). The three psychological needs associated with self-determination theory (SDT), which form the basis of Peters et al.'s [68] Motivation, Engagement, and Thriving in User Experience (METUX) model, were particularly salient in teens' descriptions of satisfying and unsatisfying technology use. Consistent with the need for *competence*, teens pointed to the satisfaction they felt after learning something new online (e.g., video editing, music arrangement) or accomplishing something challenging, such as in a video game. Many of teens' less satisfying technology experiences seemed to undermine their *autonomy*, such as through the dark pattern [59] of YouTube's auto-play that left many teens feeling as though they had wasted their time without having anything substantive to show for it. Online interactions with friends were consistently rated as highly satisfying, reflecting the importance of *relatedness* to wellbeing. These three psychological needs—competence, autonomy, and relatedness—support the *meaning* orientation to happiness and wellbeing [69]. Other orientations, such as *engagement* and *pleasure,* were also evident in teens' descriptions of their satisfying technology experiences [19,25,78]. For instance, they spoke about being absorbed in the challenge of advancing to a new level in a video game (engagement) and the enjoyment they received from watching a good movie on Netflix (pleasure).

Another contribution of the current study is the empirical insight gained from using EMAs in combination with qualitative interviews to examine teens' reactions to the coronavirus pandemic over time. In light of the pandemic's recency, our study represents one of the first examinations of common patterns and individual differences in teen wellbeing (see also [12,43]). We demonstrated that teens faced many common stressors during lockdown, such as COVID-related worry, transitioning to online school, family tensions, separation from friends, and technology over-use, as well as many of the same supports, including emotional support from friends and family, and entertainment, communication, and learning opportunities through their networked technologies. At the same time, our interviews revealed that similar stressors affected teens differently, and some teens experienced unique stressors, such as having a parent with COVID-19 or pre-existing anxiety and depression. These differences gave rise to considerable variation across teens' individual growth trajectories depicting daily fluctuations in positive feelings, sense of connection, and worry (see Figure 1, Table 2).

These insights will be useful to inform targeted interventions to support teens in response to the pandemic and during disruptive life events more generally. Teens who lack strong, supportive family relationships or friendships, and teens with pre-existing mental health challenges such as anxiety and depression appear to be most vulnerable and in need of support. Those teens who enjoy positive family relationships, have maintained strong connections with their friends, and did not experience prior mental health challenges may not need additional support. These findings align with existing knowledge of key risk and protective factors associated with adolescent wellbeing [60].

The general trend toward increased wellbeing between April and August 2020 is cause for some optimism with respect to teens' resilience amid the pandemic and can also be used to inform intervention efforts. Insights from the second interview and teens' reflections on their EMA responses over time suggest that factors contributing to teens' increased wellbeing included being able to see some of their friends in person, resumption of extracurricular activities (even if in a limited capacity), and becoming psychologically habituated to a new way of life. Thus, interventions should account for what has helped teens during the course of the pandemic and strive to incorporate, as much as possible, comparable supports.

We also present several design implications drawn from our empirical findings and participatory design sessions with teens. We intend for these implications to inform future design work exploring technology solutions to support teen wellbeing, both during disruptive life events such as the current pandemic and more generally in teens' everyday interactions with technology.

1. *Time management should emphasize purpose over productivity.* The extra time and lack of structure associated with the pandemic left teens struggling to manage their time well. They were highly aware of their increased time with technology and expressed feelings of guilt for regularly spending more time than anticipated online. Although teens expressed the desire to decrease what they perceived as wasted time with technology, prior work has shown that block-out mechanisms—a common design solution for limiting time with technology—typically frustrate users and interfere with their autonomy [88]. Therefore, we suggest that designers consider features that check in with users during the course of their technology use, inviting them to reflect on the quality of their experience and whether they are accomplishing what they set out to do. To further support a sense of autonomy, users could decide on the timing and type of check-in question/s they would like to be asked. Drawing on an idea from our design work, the check in could provide suggestions for alternate activities, whether with or without technology. This suggestion aligns with prior work that aims to support user aspirations and goal setting over punishment [31,47,48]. The overarching aim is to support users in their quest to feel that their time with technology is time well spent, including "downtime" that does not have a specific outcome or goal. This is particularly important in the context of a pandemic that increased teens' unstructured time considerably.

2. *Support the discovery and development of deeper interests.* We found that teens who had existing hobbies they could engage in during the pandemic were able to use their newfound time to pursue their interests more deeply. In contrast, those teens who lacked strong, pre-existing interests (or were unable to engage in them due to the circumstances) were more likely to feel a sense of aimlessness in their days and with their technology use. Particularly at a time when teens may not have access to their usual resources or social groups, we encourage designers to explore features that connect teens with ideas or activities they could pursue in greater depth. For instance, instead of automatically playing the next queued video, platforms like YouTube could offer users suggestions for hobbies they might pursue in relation to the video they just watched, as well as online communities that could provide them with mentorship to support and sustain their pursuits. This suggestion complements prior work that reimagines recommender systems with the goal of supporting user agency and autonomy (e.g., [26,56]).

3. *Expand and enrich teens' shrinking social worlds.* Our second design session addressed teens' feelings of loss around their diminished social worlds and their feeling that online communication lacked the richness of their face-to-face interactions with friends. Many youth were only talking to close friends that they knew well and expressed regret about being out of touch with their acquaintances. In line with prior work (e.g., [5,9,76]), the teen co-designers identified social VR as a promising technology for connecting friends—both close friends and acquaintances—around a shared, enjoyable, and meaningful experience. They felt the immersive experience of VR would make for a more satisfying interaction than video chat, and the potential to engage in a variety of different activities would give them something to talk about and share with their friends. When designing social VR experiences for teens, we suggest designers consider how to identify and connect teens based on common interests (keeping in mind safety concerns related to youth and networked technologies), the type of peer connection (close friend vs. school acquaintance vs. online-only interaction), and how to coordinate schedules to meet in a virtual world. Incorporating game-like features could further facilitate feelings of connectedness around a shared goal [34].

## 5.1 Limitations and Future Work

The current work demonstrates the importance of study designs that account for individual variation over time in teens' wellbeing and technology use, as well as the value of using complementary qualitative methods, such as semi-structured interviews, to shed light on the person-specific factors underlying longitudinal trends [51]. Despite its strengths, such an approach is time intensive, which limited the number of participants we could include in our study. Although we strove to recruit a demographically diverse group of teen participants, our sample is limited to a single geographic location in the United States. Moreover, 12 participants (57%) had at least one parent with a graduate degree, indicating a relatively high average socio-economic status (SES) in the sample. Our sample was relatively diverse with respect to race, but African Americans were under-represented. In light of early research indicating differential impacts of the pandemic across SES and race [33,43,52], future research is needed to examine whether the patterns identified in the current study are apparent in teens representing different demographic groups. We also encourage HCI researchers to use the design implications discussed here as the basis for future longer-term design work with teens around technologies that support meaning and wellbeing, both during stressful life events and in more regular, everyday life experiences.

## 6 CONCLUSION

Through interviews and ecological momentary assessments (EMAs), we show teens' diverse reactions to the coronavirus pandemic and the specific role played by technology. We identified key factors affecting teen wellbeing during the early months (April – August 2020) of the pandemic in the United States: pandemic-related worry, the transition to online school, increased family time, decreased time with friends, and teens' individual dispositions. As teens filled their newfound and newly unstructured time with networked technologies, the association between their technology use and wellbeing had more to do with the level of satisfaction they derived from their technology use than the amount of time they spent using technology. We used these empirical findings and follow-up participatory design sessions with teens to suggest a set of design implications to increase the meaningfulness of teens' technology interactions by instilling a sense of time well spent and by supporting satisfying social interactions. Although these design implications are generated from and targeted to stressful life events such as the current global pandemic, we believe they can be applied effectively to teens' technology interactions in non-crisis periods.

## ACKNOWLEDGMENTS

We would like to thank our participants for their insightful contributions and willingness to work with us for several months during a difficult time. You're amazing. Additional thanks to Kelly Meng for her assistance with interview transcription. We also thank the teachers, parents, healthcare professionals, and essential workers for their efforts.

## REFERENCES

[1] June Ahn, Tamara Clegg, Jason Yip, Elizabeth Bonsignore, Daniel Pauw, Lautaro Cabrera, Kenna Hernly, Caroline Pitt, Kelly Mills, Arturo Salazar, Diana Griffing, Jeff Rick, and Rachael Marr. 2018. Science Everywhere: Designing Public, Tangible Displays to Connect Youth Learning Across Settings. In *Proceedings of the 2018 CHI Conference on Human Factors in Computing Systems* (CHI '18), 278:1–278:12. https://doi.org/10.1145/3173574.3173852

[2] Ine Beyens, J Loes Pouwels, Irene I van Driel, Loes Keijsers, and Patti M Valkenburg. 2020. The effect of social media on well-being differs from adolescent to adolescent. *Scientific Reports* 10, 1: 1–11.

[3] Arpita Bhattacharya. 2019. Designing to Support Teen Mental Health Using Asynchronous Online Groups. In *Proceedings of the 18th ACM International Conference on Interaction Design and Children* (IDC '19), 723–727. https://doi.org/10.1145/3311927.3325352

[4] Arpita Bhattacharya, Calvin Liang, Emily Y. Zeng, Kanishk Shukla, Miguel E. R. Wong, Sean A. Munson, and Julie A. Kientz. 2019. Engaging Teenagers in Asynchronous Online Groups to Design for Stress Management. In *Proceedings of the Interaction Design and Children on ZZZ - IDC '19*, 26–37. https://doi.org/10.1145/3311927.3323140

[5] Elin A. Björling, Rachael Cicero, Aditya Sankar, and Anand Sekar. 2019. Thought Disposal: Co-Designing a virtual interaction to reduce stress in teens. In *Proceedings of the 18th ACM International Conference on Interaction Design and Children* (IDC '19), 562–567. https://doi.org/10.1145/3311927.3325313

[6] Elroy Boers, Mohammad H Afzali, Nicola Newton, and Patricia Conrod. 2019. Association of screen time and depression in adolescence. *JAMA pediatrics* 173, 9: 853–859.

[7] Richard E Boyatzis. 1998. *Transforming qualitative information: Thematic analysis and code development*. sage.

[8] Rafael A Calvo and Dorian Peters. 2014. *Positive computing: technology for wellbeing and human potential*. MIT Press.

[9] Alice Chirico, Pietro Cipresso, David B Yaden, Federica Biassoni, Giuseppe Riva, and Andrea Gaggioli. 2017. Effectiveness of immersive videos in inducing awe: an experimental study. *Scientific Reports* 7, 1: 1–11.

[10] Hui-Tzu Grace Chou and Nicholas Edge. 2012. "They are happier and having better lives than I am": the impact of using Facebook on perceptions of others' lives. *Cyberpsychology, Behavior, and Social Networking* 15, 2: 117–121.

[11] Dante Cicchetti and Fred A Rogosch. 2002. A developmental psychopathology perspective on adolescence. *Journal of consulting and clinical psychology* 70, 1: 6.

[12] Common Sense Media and Survey Monkey. 2020. *How teens are coping and connecting in the time of the coronavirus*. Retrieved from https://www.commonsensemedia.org/sites/default/files/uploads/pdfs/2020_surveymonkey-key-findings-toplines-teens-and-coronavirus.pdf

[13] Bruce E Compas, Pamela G Orosan, and Kathryn E Grant. 1993. Adolescent stress and coping: Implications for psychopathology during adolescence. *Journal of adolescence* 16, 3: 331.

[14] Darren Courtney, Priya Watson, Marco Battaglia, Benoit H Mulsant, and Peter Szatmari. 2020. COVID-19 impacts on child and youth anxiety and depression: challenges and opportunities. *The Canadian Journal of Psychiatry*: 0706743720935646.

[15] David Coyle, Gavin Doherty, and John Sharry. 2010. PlayWrite: end-user adaptable games to support adolescent mental health. In *CHI'10 Extended Abstracts on Human Factors in Computing Systems*. 3889–3894.

[16] David Coyle, Nicola McGlade, Gavin Doherty, and Gary O'Reilly. 2011. Exploratory evaluations of a computer game supporting cognitive behavioural therapy for adolescents. In *Proceedings of the SIGCHI Conference on Human Factors in Computing Systems*, 2937–2946.

[17] Sarah M Coyne, Laura M Padilla-Walker, and Emily Howard. 2013. Emerging in a digital world: A decade review of media use, effects, and gratifications in emerging adulthood. *Emerging Adulthood* 1, 2: 125–137.

[18] Sarah M Coyne, Adam A Rogers, Jessica D Zurcher, Laura Stockdale, and McCall Booth. 2020. Does time spent using social media impact mental health?: An eight year longitudinal study. *Computers in Human Behavior* 104: 106160.

[19] Mihaly Csikszentmihalyi and Mihaly Csikzentmihaly. 1990. *Flow: The psychology of optimal experience.* Harper & Row New York.

[20] William Damon and Heather Malin. 2020. The Development of Purpose. *The Oxford Handbook of Moral Development: An Interdisciplinary Perspective*: 110.

[21] Katie Davis. 2013. Young people's digital lives: The impact of interpersonal relationships and digital media use on adolescents' sense of identity. *Computers in Human Behavior* 29, 6: 2281–2293.

[22] Edward L Deci and Richard M Ryan. 2000. The" what" and" why" of goal pursuits: Human needs and the self-determination of behavior. *Psychological inquiry* 11, 4: 227–268.

[23] Pieter MA Desmet and Anna E Pohlmeyer. 2013. Positive design: An introduction to design for subjective well-being. *International journal of design* 7, 3.

[24] ED Diener, Robert A Emmons, Randy J Larsen, and Sharon Griffin. 1985. The satisfaction with life scale. *Journal of personality assessment* 49, 1: 71–75.

[25] Ed Diener and Martin EP Seligman. 2002. Very happy people. *Psychological science* 13, 1: 81–84.

[26] Daniella DiPaola, Blakeley H Payne, and Cynthia Breazeal. 2020. Decoding design agendas: an ethical design activity for middle school students. In *Proceedings of the Interaction Design and Children Conference*, 1–10.

[27] Jerry Alan Fails, Mona Leigh Guha, and Allison Druin. 2013. Methods and Techniques for Involving Children in the Design of New Technology for Children. *Foundations and Trends® in Human–Computer Interaction* 6, 2: 85–166. https://doi.org/10.1561/1100000018

[28] Ezra Golberstein, Hefei Wen, and Benjamin F Miller. 2020. Coronavirus disease 2019 (COVID-19) and mental health for children and adolescents. *JAMA pediatrics*.

[29] Marc Hassenzahl. 2010. Experience design: Technology for all the right reasons. *Synthesis lectures on human-centered informatics* 3, 1: 1–95.

[30] Patrick L Hill, Nancy L Sin, Nicholas A Turiano, Anthony L Burrow, and David M Almeida. 2018. Sense of purpose moderates the associations between daily stressors and daily well-being. *Annals of Behavioral Medicine* 52, 8: 724–729.

[31] Alexis Hiniker, Sungsoo Hong, Tadayoshi Kohno, and Julie A Kientz. 2016. MyTime: designing and evaluating an intervention for smartphone non-use. In *Proceedings of the 2016 CHI Conference on Human Factors in Computing Systems*, 4746–4757.

[32] Alexis Hiniker, Jenny S Radesky, Sonia Livingstone, and Alicia Blum-Ross. 2019. Moving Beyond" The Great Screen Time Debate" in the Design of Technology for Children. In *Extended Abstracts of the 2019 CHI Conference on Human Factors in Computing Systems*, 1–6.

[33] Monica Webb Hooper, Anna María Nápoles, and Eliseo J Pérez-Stable. 2020. COVID-19 and racial/ethnic disparities. *Jama*.

[34] Ioanna Iacovides and Elisa D Mekler. 2019. The Role of Gaming During Difficult Life Experiences. In *Proceedings of the 2019 CHI Conference on Human Factors in Computing Systems*, 1–12.

[35] Mizuko Ito, Candice Odgers, Stephen Schueller, Jennifer Cabrera, Evan Conaway, Remy Cross, and Maya Hernandez. 2020. *Social Media and Youth Wellbeing: What We Know and Where We Could Go.* Connected Learning Alliance, Irvine, CA.

[36] Michaeline Jensen, Madeleine J George, Michael R Russell, and Candice L Odgers. 2019. Young adolescents' digital technology use and mental health symptoms: Little evidence of longitudinal or daily linkages. *Clinical Psychological Science* 7, 6: 1416–1433.

[37] Minsam Ko, Subin Yang, Joonwon Lee, Christian Heizmann, Jinyoung Jeong, Uichin Lee, Daehee Shin, Koji Yatani, Junehwa Song, and Kyong-Mee Chung. 2015. NUGU: a group-based intervention app for improving self-regulation of limiting smartphone use. In *Proceedings of the 18th ACM conference on computer supported cooperative work & social computing*, 1235–1245.

[38] Ethan Kross, Philippe Verduyn, Emre Demiralp, Jiyoung Park, David Seungjae Lee, Natalie Lin, Holly Shablack, John Jonides, and Oscar Ybarra. 2013. Facebook use predicts declines in subjective well-being in young adults. *PloS one* 8, 8: e69841.

[39] Nan M Laird and James H Ware. 1982. Random-effects models for longitudinal data. *Biometrics*: 963–974.

[40] Simone Lanette, Phoebe K Chua, Gillian Hayes, and Melissa Mazmanian. 2018. How Much is' Too Much'? The Role of a Smartphone Addiction Narrative in Individuals' Experience of Use. *Proceedings of the ACM on Human-Computer Interaction* 2, CSCW: 1–22.

[41] Alexis Lauricella R., Drew Cingel P., Leanne Beaudoin-Ryan, Micheal B Robb, Melissa Saphir, and Ellen Wartella. 2016. *The Common Sense census: Plugged-in parents of tweens and teens.* Common Sense Media, San Francisco, CA.

[42] Heidi M Levitt, Michael Bamberg, John W Creswell, David M Frost, Ruthellen Josselson, and Carola Suárez-Orozco. 2018. Journal article reporting standards for qualitative primary, qualitative meta-analytic, and mixed methods research in psychology: The APA Publications and Communications

Board task force report. *American Psychologist* 73, 1: 26.

[43] Leilei Liang, Hui Ren, Ruilin Cao, Yueyang Hu, Zeying Qin, Chuanen Li, and Songli Mei. 2020. The effect of COVID-19 on youth mental health. *Psychiatric Quarterly*: 1–12.

[44] Alice J Lin, Charles B Chen, and Fuhua Frank Cheng. 2019. Multiplayer Virtual Reality Game for Pediatric Patients. In *Proceedings of the 2019 4th International Conference on Multimedia Systems and Signal Processing*, 47–51.

[45] Leslie S Liu, Kori M Inkpen, and Wanda Pratt. 2015. " I'm Not Like My Friends" Understanding How Children with a Chronic Illness Use Technology to Maintain Normalcy. In *Proceedings of the 18th ACM Conference on Computer Supported Cooperative Work & Social Computing*, 1527–1539.

[46] Kai Lukoff, Cissy Yu, Julie Kientz, and Alexis Hiniker. 2018. What Makes Smartphone Use Meaningful or Meaningless? *Proceedings of the ACM on Interactive, Mobile, Wearable and Ubiquitous Technologies* 2, 1: 1–26.

[47] Ulrik Lyngs, Kai Lukoff, Petr Slovak, Reuben Binns, Adam Slack, Michael Inzlicht, Max Van Kleek, and Nigel Shadbolt. 2019. Self-Control in Cyberspace: Applying Dual Systems Theory to a Review of Digital Self-Control Tools. In *Proceedings of the 2019 CHI Conference on Human Factors in Computing Systems*, 1–18.

[48] Ulrik Lyngs, Kai Lukoff, Petr Slovak, William Seymour, Helena Webb, Marina Jirotka, Jun Zhao, Max Van Kleek, and Nigel Shadbolt. 2020. "I Just Want to Hack Myself to Not Get Distracted" Evaluating Design Interventions for Self-Control on Facebook. In *Proceedings of the 2020 CHI Conference on Human Factors in Computing Systems*, 1–15.

[49] Dorit Maor and Katherine Mitchem. 2020. Hospitalized adolescents' use of mobile technologies for learning, communication, and Well-being. *Journal of Adolescent Research* 35, 2: 225–247.

[50] Michael Massimi, Jill P Dimond, and Christopher A Le Dantec. 2012. Finding a new normal: the role of technology in life disruptions. In *Proceedings of the acm 2012 conference on computer supported cooperative work*, 719–728.

[51] Joseph A Maxwell. 2012. *Qualitative research design: An interactive approach*. Sage publications.

[52] Eoin McElroy, Praveetha Patalay, Bettina Moltrecht, Mark Shevlin, Adrienne Shum, Cathy Creswell, and Polly Waite. 2020. Demographic and health factors associated with pandemic anxiety in the context of COVID-19.

[53] Institute of Digital Media, Child Development Working Group on Games for Health, Tom Baranowski, Fran Blumberg, Richard Buday, Ann DeSmet, Lynn E Fiellin, C Shawn Green, Pamela M Kato, Amy Shirong Lu, Ann E Maloney, and others. 2016. Games for health for children—Current status and needed research. *Games for health journal* 5, 1: 1–12.

[54] Elisa D Mekler and Kasper Hornbæk. 2019. A framework for the experience of meaning in human-computer interaction. In *Proceedings of the 2019 CHI Conference on Human Factors in Computing Systems*, 1–15.

[55] Mark Merolli, Kathleen Gray, Fernando Martin-Sanchez, and Guillermo Lopez-Campos. 2015. Patient-reported outcomes and therapeutic affordances of social media: findings from a global online survey of people with chronic pain. *Journal of medical Internet research* 17, 1: e20.

[56] Simon Meyffret, Lionel Médini, and Frédérique Laforest. 2012. Trust-based local and social recommendation. In *Proceedings of the 4th ACM RecSys workshop on Recommender systems and the social web*, 53–60.

[57] Matthew B Miles and A Michael Huberman. 1994. *Qualitative data analysis: An expanded sourcebook*. sage.

[58] Ramin Mojtabai, Mark Olfson, and Beth Han. 2016. National trends in the prevalence and treatment of depression in adolescents and young adults. *Pediatrics* 138, 6: e20161878.

[59] Arvind Narayanan, Arunesh Mathur, Marshini Chetty, and Mihir Kshirsagar. 2020. Dark Patterns: Past, Present, and Future. *Queue* 18, 2: 67–92.

[60] National Academies of Sciences Engineering, Medicine, and others. 2020. *Promoting positive adolescent health behaviors and outcomes: Thriving in the 21st century*. National Academies Press.

[61] Candice L Odgers and Michaeline R Jensen. 2020. Annual Research Review: Adolescent mental health in the digital age: facts, fears, and future directions. *Journal of Child Psychology and Psychiatry* 61, 3: 336–348.

[62] Yoon Phaik Ooi, Dion Hoe-Lian Goh, Elisa D Mekler, Alexandre N Tuch, Jillian Boon, Rebecca P Ang, Daniel Fung, and Jens Gaab. 2016. Understanding player perceptions of RegnaTales, a mobile game for teaching social problem solving skills. In *Proceedings of the 31st Annual ACM Symposium on Applied Computing*, 167–172.

[63] Amy Orben. 2020. Teenagers, screens and social media: a narrative review of reviews and key studies. *Social psychiatry and psychiatric epidemiology*: 1–8.

[64] Amy Orben, Tobias Dienlin, and Andrew K Przybylski. 2019. Social media's enduring effect on adolescent life satisfaction. *Proceedings of the National Academy of Sciences* 116, 21: 10226–10228.

[65] Amy Orben and Andrew K Przybylski. 2019. The association between adolescent well-being and digital technology use. *Nature Human Behaviour* 3, 2: 173–182.

[66] W Pavot and E Diener. 2013. *The Satisfaction with Life Scale (SWL). Measurement instrument database for the social science*.

[67] Samantha Pearcey, Adrienne Shum, Polly Waite, Praveetha Patalay, and Cathy Creswell. 2020. *Report 04: Changes in children and young people's emotional and behavioral difficulties through lockdown*. University of Oxford, Oxford, UK. Retrieved September 7, 2020 from http://localhost/minervation/co-space/findings/

[68] Dorian Peters, Rafael A Calvo, and Richard M Ryan. 2018. Designing for motivation, engagement and wellbeing in digital experience. *Frontiers in psychology* 9: 797.

[69] Christopher Peterson, Nansook Park, and Martin EP Seligman. 2005. Orientations to happiness and life satisfaction: The full life versus the empty life. *Journal of happiness studies* 6, 1: 25–41.

[70] Sarah Pink, Heather Horst, John Postill, Larissa Hjorth, Tania Lewis, and Jo Tacchi. 2015. *Digital Ethnography: Principles and Practice*. SAGE.

[71] Emmet Power, Sarah Hughes, David Cotter, and Mary Cannon. 2020. Youth Mental Health in the time of COVID-19. *Irish Journal of Psychological Medicine*: 1–15.

[72] Jenny S Radesky, Jayna Schumacher, and Barry Zuckerman. 2015. Mobile and interactive media use by young children: the good, the bad, and the unknown. *Pediatrics* 135, 1: 1–3.

[73] Nilam Ram, Xiao Yang, Mu-Jung Cho, Miriam Brinberg, Fiona Muirhead, Byron Reeves, and Thomas N Robinson. 2020. Screenomics: A new approach for observing and studying individuals' digital lives. *Journal of Adolescent Research* 35, 1: 16–50.

[74] Victoria Rideout and Micheal B Robb. 2018. Social media, social life: Teens reveal their experiences. *San Francisco, CA: Common Sense Media*.

[75] Victoria Rideout and Micheal B Robb. 2019. *The Common Sense census: Media use by tweens and teens, 2019*. Common Sense Media, San Francisco, CA.

[76] Giuseppe Riva, Fabrizia Mantovani, and Brenda K Wiederhold. 2020. Positive Technology and COVID-19. *Cyberpsychology, Behavior, and Social Networking*.

[77] Richard M Ryan and Edward L Deci. 2020. Intrinsic and extrinsic motivation from a self-determination theory perspective: Definitions, theory, practices, and future directions. *Contemporary Educational Psychology*: 101860.

[78] Stephen M Schueller and Martin EP Seligman. 2010. Pursuit of pleasure, engagement, and meaning: Relationships to subjective and objective measures of well-being. *The Journal of Positive Psychology* 5, 4: 253–263.

[79] Martin EP Seligman. 2012. *Flourish: A visionary new understanding of happiness and well-being*. Simon and Schuster.

[80] Saul Shiffman, Arthur A Stone, and Michael R Hufford. 2008. Ecological momentary assessment. *Annu. Rev. Clin. Psychol.* 4: 1–32.

[81] Judith D Singer, John B Willett, John B Willett, and others. 2003. *Applied longitudinal data analysis: Modeling change and event occurrence*. Oxford university press.

[82] Peter Smagorinsky. 2008. The method section as conceptual epicenter in constructing social science research reports. *Written communication* 25, 3: 389–411.

[83] Rachel C. Smith, Ole S. Iversen, Thomas Hjermitslev, and Aviaja B. Lynggaard. 2013. Towards an ecological inquiry in child-computer interaction. In *Proceedings of the 12th International Conference on Interaction Design and Children* (IDC '13), 183–192. https://doi.org/10.1145/2485760.2485780

[84] Leah H Somerville. 2013. The teenage brain: Sensitivity to social evaluation. *Current directions in psychological science* 22, 2: 121–127.

[85] Laurence Steinberg and Amanda Sheffield Morris. 2001. Adolescent development. *Annual review of psychology* 52, 1: 83–110.

[86] Andrew Steptoe and Jane Wardle. 2011. Positive affect measured using ecological momentary assessment and survival in older men and women. *Proceedings of the National Academy of Sciences* 108, 45: 18244–18248.

[87] Katie Headrick Taylor, Lori Takeuchi, and Reed Stevens. 2018. Mapping the daily media round: novel methods for understanding families' mobile technology use. *Learning, Media and Technology* 43, 1: 70–84. https://doi.org/10.1080/17439884.2017.1391286

[88] Jonathan A Tran, Katie S Yang, Katie Davis, and Alexis Hiniker. 2019. Modeling the engagement-disengagement cycle of compulsive phone use. In *Proceedings of the 2019 CHI Conference on Human Factors in Computing Systems*, 1–14.

[89] Jean M Twenge, Thomas E Joiner, Megan L Rogers, and Gabrielle N Martin. 2018. Increases in depressive symptoms, suicide-related outcomes, and suicide rates among US adolescents after 2010 and links to increased new media screen time. *Clinical Psychological Science* 6, 1: 3–17.

[90] Jean M Twenge, Gabrielle N Martin, and W Keith Campbell. 2018. Decreases in psychological well-being among American adolescents after 2012 and links to screen time during the rise of smartphone technology. *Emotion* 18, 6: 765.

[91] UK Youth. 2020. *The impact of COVID-19 on young people and the youth sector*. UK Youth, London. Retrieved September 7, 2020 from https://www.ukyouth.org/wp-content/uploads/2020/04/UK-Youth-Covid-19-Impact-Report-External-Final-08.04.20.pdf

[92] Patti M Valkenburg and Jochen Peter. 2013. The differential susceptibility to media effects model. *Journal of communication* 63, 2: 221–243.

[93] Philippe Verduyn, David Seungjae Lee, Jiyoung Park, Holly Shablack, Ariana Orvell, Joseph Bayer, Oscar Ybarra, John Jonides, and Ethan Kross. 2015. Passive Facebook usage undermines affective well-being: Experimental and longitudinal evidence. *Journal of Experimental Psychology: General* 144, 2: 480.

[94] Greg Walsh, Allison Druin, Mona Leigh Guha, Elizabeth Bonsignore, Elizabeth Foss, Jason C. Yip, Evan Golub, Tamara Clegg, Quincy Brown, and Robin Brewer. 2012. DisCo: a co-design online tool for asynchronous distributed child and adult design partners. In *Proceedings of the 11th International Conference on Interaction Design and Children*, 11–19.

[95] Greg Walsh, Elizabeth Foss, Jason Yip, and Allison Druin. 2013. FACIT PD: A Framework for Analysis and Creation of Intergenerational Techniques for Participatory Design. In *Proceedings of the SIGCHI Conference on Human Factors in Computing Systems* (CHI '13), 2893–2902. https://doi.org/10.1145/2470654.2481400

[96] World Health Organization. 2019. *Coming of age: Adolescent health*. Retrieved from https://www.who.int/news-room/spotlight/coming-of-age-adolescent-health

[97] Jason C. Yip, Kiley Sobel, Caroline Pitt, Kung Jin Lee, Sijin Chen, Kari Nasu, and Laura R. Pina. 2017. Examining Adult-Child Interactions in Intergenerational Participatory Design. In *Proceedings of the 2017 CHI Conference on Human Factors in Computing Systems* (CHI '17), 5742–5754. https://doi.org/10.1145/3025453.3025787

[98] Childstats.gov - America's Children: Key National Indicators of Well-Being, 2019 - Adolescent Depression. Retrieved September 6, 2020 from https://www.childstats.gov/americaschildren/health4.asp